\documentclass[a4paper,fleqn,usenatbib]{mnras}

\usepackage{newtxtext,newtxmath}
\usepackage[T1]{fontenc}
\usepackage{ae,aecompl}
\usepackage{graphicx}	% Including figure files
\usepackage{amsmath}	% Advanced maths commands
\usepackage{amssymb}	% Extra maths symbols
\usepackage{bm}

\newcommand{\bfx}{\mathbfit{x}}
\newcommand{\bfTheta}{\mathbf{\Theta}}
\newcommand{\bftheta}{\bm{\theta}}
\newcommand{\bflambda}{\bm{\lambda}}
\newcommand{\bfmu}{\bm{\mu}}
\newcommand{\bfv}{\mathbfit{v}}
\newcommand{\bfJ}{\mathbfit{J}}
\newcommand{\bfj}{\mathbfit{j}}
\newcommand{\bnabla}{\mathbf{\nabla}}
\newcommand{\half}{\textstyle{\frac{1}{2}}}
\newcommand{\bfP}{\mathbfit{P}}

\newcommand{\matS}{\mathbfss{S}}
\newcommand{\matT}{\mathbfss{T}}
\newcommand{\p}{\upartial}
\newcommand{\pr}{\mbox{Pr}\,}
\newcommand{\ud}{{\rm d}}

\newfont{\caps}{cmcsc10}
\newcommand{\df}{{\caps df}}

\title[Masses of stellar systems]{Measuring the mass distribution in stellar systems}

\author[S.\ Tremaine]{
Scott Tremaine$^{1}$\thanks{E-mail: tremaine@ias.edu}
\\
$^{1}$Institute for Advanced Study, Einstein Drive, Princeton, NJ
08540, USA}

\pubyear{2017}

\begin{document}
\label{firstpage}
\pagerange{\pageref{firstpage}--\pageref{lastpage}}
\maketitle

\begin{abstract}
  One of the fundamental tasks of dynamical astronomy is to infer the
  distribution of mass in a stellar system from a snapshot of the
  positions and velocities of its stars. The usual approach to this
  task (e.g., Schwarzschild's method) involves fitting parametrized
  forms of the gravitational potential and the phase-space
  distribution to the data. We review the practical and conceptual
  difficulties with this approach and describe a novel statistical
  method for determining the mass distribution that does not require
  determining the phase-space distribution of the stars. We show that this new
  estimator out-performs other distribution-free estimators for the
  harmonic and Kepler potentials.
 \end{abstract}

\begin{keywords}
stars: kinematics and dynamics -- galaxies: kinematics and dynamics -- methods: statistical 
\end{keywords}

\section{Introduction}

\label{sec:intro}

\noindent
The determination of the mass distribution in stellar systems has led
to some of the most important discoveries in astronomy. Examples
include Kapteyn's (1922) \nocite{kap22} measurement of the local
density in the Galactic disc; Zwicky's (1933) \nocite{zw33} discovery
that the dark mass in the Coma cluster of galaxies is several hundred
times larger than the mass in stars; measurements of the mass
distribution in low-luminosity dwarf galaxies which show them also to
be dominated by dark matter \citep[e.g.,][]{aar1983,mac12}; and the
determination of the mass of the central black hole in our Galaxy from
the kinematics of the nuclear star cluster
\citep[e.g.,][]{saha01,feld16}. The scope and power of such
measurements will grow in the near future, when the {\it Gaia}
spacecraft provides accurate positions and velocities for $\sim 10^9$
stars in the Galaxy.

In this paper we examine an idealized formulation of this task. A set
of $N$ test particles orbit in a gravitational potential
$\Phi(\bfx|\,\bfmu)$, where $\bfmu=(\mu_1,\mu_2,\ldots)$ is a list of
parameters that determine the analytic form of the potential. We call
these mass parameters since the mass distribution follows directly
from the potential via Poisson's equation $\nabla^2\Phi=4\upi
G\rho$. The system is assumed to be in a steady state, that is, the
particles are distributed randomly in orbital phase. We know the
positions and velocities of the particles at some instant,
$\{\bfx_n,\bfv_n\},$, $n=1,\ldots,N$. What can we infer about the mass
parameters $\bfmu$?

Perhaps the simplest astronomical example of this problem is to
estimate the force law in the solar system from a snapshot of the
positions and velocities of the eight planets. \cite{bovy+2012} use
this example to motivate a wide-ranging discussion of the application
of Bayesian methods to dynamical inference. 

The idealized problem we examine here has only limited relevance to
real astronomical systems, for several reasons. In most systems only
two of the three spatial coordinates are known because the distance
along the line of sight cannot be determined, and only one of the
three velocity components is known because the proper motions are
undetectably small. In addition, the survey region is often smaller
than the system, so only a subset of the stars is measured. Finally,
the measurement uncertainties in the velocities are often
substantial. We shall not discuss these important practical issues, in
order to focus on the simplest version of the task of mass
determination, which is already complicated enough. 

The test particles sample a distribution function (hereafter \df)
$F(\bfx,\bfv)$, defined such that $F(\bfx,\bfv)\, d\bfx d\bfv$ is the
probability that a given particle is found in the small phase-space volume
$d\bfx d\bfv$.  Thus $\int F(\bfx,\bfv)\,d\bfx d\bfv=1$. Since the
system is in a steady state, $F(\bfx,\bfv)$ can only depend on the
integrals of motion in the potential $\Phi(\bfx|\,\bfmu)$ (Jeans's
theorem). Otherwise $F$ is arbitrary so long as it is non-negative.
Thus, the observations are consistent with a given set of mass
parameters $\bfmu$ if and only if the inferred density of the test particles is
constant and non-negative on phase-space surfaces on which the
integrals of motion are fixed. 

For simplicity we assume that motion in the potential
$\Phi(\bfx|\,\bfmu)$ is integrable. Then if the phase space has $D$
degrees of freedom ($2D$ dimensions) there exist $D$ action-angle
pairs $(\bfj,\bftheta)$ that form canonical momenta and
coordinates for each particle. The relation between the Cartesian coordinates
$(\bfx,\bfv)$ and the actions and angles is given by the functions
$\bfj=\bfJ(\bfx,\bfv|\,\bfmu)$ and
$\bftheta=\bfTheta(\bfx,\bfv|\,\bfmu)$. Since the actions are integrals of
motion, we can write the \df\ as $F(\bfj)$ with
$\int F(\bfj) \,\ud\bfj \ud\bftheta = (2\upi)^D\!\int F(\bfj) \,\ud\bfj=1$.

\subsection{Schwarzschild's method}

The usual approach to estimating the mass parameters $\bfmu$ starts by
assuming that the \df\ depends on a set of parameters
$\bflambda=(\lambda_1,\lambda_2,\ldots,\lambda_L)$, and then infers
both $\bfmu$ and $\bflambda$ from the data. Since we are not
interested in the properties of the \df\ (for the purposes of this paper),
$\{\lambda_l\}$ are called nuisance parameters. In practice the \df\ is
assumed to have the form
\begin{equation}
F(\bfj|\,\bflambda) = \sum_{l=1}^L\lambda_l \,\psi_l(\bfj ), 
\end{equation} 
where $\{\psi_l(\bfj)\}$ is a set of basis functions in action space
and $\{\lambda_l\}$ is a set of weights subject to the constraint that
$F(\bfj|\,\bflambda)$ cannot be negative.  In the simplest and most
widely used variant, Schwarzschild's method, the basis functions are
individual orbits,
\begin{equation}
\psi_l(\bfj)=\delta(\bfj-\bfj_l), 
\end{equation}
where $\{\bfj_l\}$ is an orbit `library' that covers the action space as
well as possible. Most users of Schwarzschild's method then maximize
the likelihood of the data over the parameters $\bfmu$ and
$\bflambda$. This approach raises a number of concerns:

\begin{enumerate}

\item The dimensionality of $\bflambda$ is generally far larger than
  the dimensionality of $\bfmu$. Thus the data are being used mainly
  to determine nuisance parameters rather than the mass distribution.

\item The number of nuisance parameters describing the \df, $L$,  is
  likely to be comparable to, or even exceed, the number of particles $N$. In
  this case maximum likelihood can give inconsistent results; in other
  words, when $N\sim L$ the maximum-likelihood estimate
  $\hat\bfmu$ may not converge to the correct value $\bfmu$ as
  $N\to\infty$  \citep[e.g.,][]{ns48,lan2000}.

\item An alternative is to use Bayesian rather than
  maximum-likelihood methods. The Bayesian approach is to marginalize
  over all of the nuisance parameters $\bflambda$ using some suitable
  priors. Thus the posterior probability of the mass parameters
  $\bfmu$ is
\begin{equation}
p(\bfmu|\, \{\bfx_n,\bfv_n\}) = C\,
\pr(\bfmu) \int \ud\bflambda\, \pr(\bflambda)\prod_{n=1}^N
F[\bfJ(\bfx_n,\bfv_n|\,\bfmu)|\,\bflambda].
\label{eq:one}
\end{equation}
Here $\pr(\bfmu)$ and $\pr(\bflambda)$ are the prior probabilities of
the parameters, and $C$ is a normalization constant defined such that
$\int \ud\bfmu\,p(\bfmu|\,\{\bfx_n,\bfv_n\})=1$.  \cite{mag06} argues
that in Schwarzschild's method the prior $\pr(\bflambda)$ should be
independent of the partition of phase space (the choice of the orbit
library) and this requires that the priors are infinitely
divisible. The uniform and Jeffreys priors,
$\pr(\lambda)\propto\mbox{\,const}$ and $1/\lambda$ respectively, do
not have this property, but Magorrian describes how to find priors
that do. Unfortunately, even with infinitely divisible priors the
results depend on the choice of prior. Moreover this approach is far
more computationally expensive than maximum likelihood.

\item It is straightforward to generalize the Bayesian Schwarzschild's
  method to account for observational errors in the positions and
  velocities. Magorrian's (2014) paradox \nocite{mag14} arises when
  the orbit library is much larger than the number of particles. Then
  as the observational errors shrink to zero the mass parameters
  become less and less accurate, and in the limit where there are no
  errors there are {\it no} posterior constraints on the mass
  parameters. The reason is that the region of phase space
  corresponding to the orbit $\bfj_l$ always has either zero or one
  particle in it, whatever the mass parameters may be.

\end{enumerate}

\noindent
Schwarzschild's method has successfully passed many tests on mock
data, so it remains unclear whether these conceptual
concerns affect the many important results that have been derived from
real data using this approach. 

\subsection{Distribution-free methods} 

By `distribution-free' we mean, loosely, that the estimate of the
mass parameters does not rely on assumptions about the \df\ of the
system\footnote{This is a weaker definition than the one often used in
  statistics, in which {\it all} statistical properties of a distribution-free estimator are
  independent of the underlying distribution.}. The simplest example
of a distribution-free method is the virial theorem: if there is a
single mass parameter $\mu$ and the potential is $\Phi(\bfx|\,\mu)$,
then an estimator of the mass, $\hat \mu$, is given implicitly by
\begin{equation}
\sum_{n=1}^N v_n^2=\sum_{n=1}^N
\bfx\cdot\bnabla\Phi(\bfx|\,\hat\mu)|_{\bfx=\bfx_n}.
\label{eq:virial}
\end{equation}
For example, in the harmonic potential $\Phi(x|\omega)=\half\omega^2 x^2$ the
virial-theorem estimator takes the form
\begin{equation}
\hat\omega=\left(\frac{\sum_n v_n^2}{\sum_n x_n^2}\right)^{1/2}.
\label{eq:virial-h}
\end{equation}
The variance in $\hat\omega$ for $N\gg1$ is
\begin{equation}
\sigma^2=\frac{\omega^2}{2N}\frac{\langle j^2\rangle}{\langle
j\rangle^2},
\label{eq:sho-var}
\end{equation}
where $\langle\cdot\rangle$ denotes an average over phase space and
$j$ is the action (eq.\ \ref{eq:t2}).
For the Kepler potential $\Phi(\bfx|\mu)=-\mu/r$, 
\begin{equation}
\hat\mu=\frac{\sum_{n=1}^N v_n^2}{\phantom{\big|}\sum_{n=1}^N |\bfx_n|^{-1}}.
\label{eq:virial-m} 
\end{equation}

Unfortunately virial-theorem estimators have several undesirable
properties \citep[e.g.,][]{bt81,bl04}:

\begin{enumerate}

\item  The estimator is generally biased, that is, 
$\langle\hat\mu\rangle_\theta\not=\mu$ where $\langle\cdot\rangle_\theta$
denotes an average over the angles corresponding to the true mass. The
bias is typically $\sim\mu/N$. 

\item The estimator cannot be applied to systems in which
  there is more than one mass parameter.

\item The estimator is inefficient. For example, if the particles in a
  harmonic potential have a wide range of actions, the sums
  in both the numerator and denominator of equation
  (\ref{eq:virial-h}) are usually dominated by the particles with the
  largest or smallest action, depending on the \df. A more formal
  argument is that the ratio $\langle j^2\rangle/\langle j\rangle^2$ in
  the expression (\ref{eq:sho-var}) for the variance always exceeds
  unity by the Cauchy--Schwarz inequality, and can be much larger than
  unity if a wide range of actions is present. Thus the kinematic
  information in most of the particles is diluted. 

\item The estimator can be inconsistent, that is, it may not converge
  in probability to the correct value of $\mu$ as $N\to\infty$. In a
  harmonic potential, this problem arises when the number density of
  particles satisfies $dn \propto j^{-b}dj$ as $j\to\infty$ with $1<b\le 3$. Then
  $\langle j^2\rangle$ diverges and the sums in
  (\ref{eq:virial-h}) are dominated by a few particles with the
  largest actions, no matter how many particles are present. A similar
  problem occurs in the Kepler potential when the number density of
  particles in semimajor axis satisfies $dn\propto a^{-c} da$ as
  $a\to 0$ with $0\le c<1$. 

\end{enumerate}

\noindent
A second distribution-free method is orbital roulette \citep{bl04},
which constrains $\bfmu$ by requiring that the distribution of angles
$\bfTheta(\bfx_n,\bfv_n|\bfmu)$ must be consistent with a uniform
distribution. However this requirement is a method for hypothesis
testing rather than parameter estimation, that is, it does not specify
the distribution of angles that would obtain if the mass parameter
were different from the one assumed. Applying hypothesis testing to a
parameter-estimation problem such as this one can be misleading or
inefficient. For example:

\begin{enumerate}

\item One version of roulette described by \cite{bl04} is the `mean
  orbital phase' estimator. For any trial value of the mass
  parameter, $\mu_t$, the corresponding angle variable $\theta_t$
  (chosen so $\theta_t=0$ at periapsis) is
  folded and rescaled to a new variable
\begin{equation}
g_t =\frac{1}{\upi}\left\{\begin{array}{ll} \theta_t, &
    0\le\theta_t \le \upi, \\ 2\upi-\theta_t, & \upi <
    \theta_t < 2\upi.
\end{array}\right.
\end{equation}
The mean phase is then computed as
\begin{equation}
\overline g_t\equiv \frac{1}{N}\sum_{n=1}^N g_{t,n}.
\label{eq:uniform}
\end{equation}
If the trial mass $\mu_t$ equals the true mass $\mu$, then $\langle\,
\overline g_t\rangle_\theta=\half$, where $\langle\cdot\rangle_\theta$ denotes the
average over the angle variable $\theta$ corresponding to the true
mass.  The estimated mass $\hat\mu$ is the mass at which $\overline
g_t=\half$ and its standard deviation $\sigma$ is defined such that
over the domain $\hat\mu\pm\sigma$ the range of $\overline g_t$ is
$\half\pm (12N)^{-1/2}$.  \cite{bl04} show that this estimator works
well for the Kepler potential, but when applied to the
harmonic potential it fails dramatically, since the first of
equations (\ref{eq:sho}) below shows that $\langle\, \overline
g_t\rangle_\theta=\half$ for {\it all} values of the trial frequency
$\omega_t$. Thus the mean orbital phase estimator is useless in this case. 

\item Other versions of roulette are based on statistics such as the
  Kolmogorov--Smirnov or Anderson--Darling tests of uniformity. The
  K-S statistic $D=\max_g|F_N(g)-g|$ where $F_N(g)$ is the cumulative
  distribution of the trial phases $\{ g_{t,n}\}$; and if
  $D>D_{90}(N)$ (where for example $D_{90}(100)=0.1208$), then the
  uniform hypothesis can be rejected at the 90 per cent level. \cite{bl04}
  argue that if $D>D_{90}(N)$ for all trial masses less than
  $\mu_\mathrm{min}$ or greater than $\mu_\mathrm{max}$, then the true
  mass parameter lies in the interval
  $(\mu_\mathrm{min},\mu_\mathrm{max})$ at the 90 per cent confidence
  level. However, in this approach (i) there will be {\it no}
  acceptable mass parameter in a significant fraction of cases; (ii) in
  some cases the minimum value of the KS statistic over all trial
  masses will be only slightly less than $D_{90}(N)$, so
  $\mu_\mathrm{min}$ and $\mu_\mathrm{max}$ we be very close, leading
  to unrealistically small error bars.

\end{enumerate}

\noindent
In the following sections we describe a distribution-free mass
estimator for an arbitrary integrable potential
$\Phi(\bfx|\,\bfmu)$. We call this the `generating-function' 
estimator since it is based on the infinitesimal generating function
that relates action-angle variables at nearby values of the mass
parameters. The generating-function estimator is derived using an iterative approach:
we assume a trial value $\bfmu_t$ for the mass parameters, derive an
estimator $\hat\bfmu$, then replace $\bfmu_t$ by $\hat\bfmu$ and
iterate to convergence. In the potentials we have examined, our estimator
is consistent, that is, for all {\df}s the estimator $\hat\bfmu$ converges in probability to the
true mass parameters $\bfmu$ as $N\to\infty$. The estimator is also
unbiased in the following sense: if the trial mass parameter is equal
to the true value $\mu$, then $\langle \hat\bfmu\rangle_\theta$ is
also equal to $\bfmu$, for any sample size $N$. Here
$\langle\cdot\rangle_\theta$ denotes an average over the angle
variables corresponding to the true mass parameter.

The estimator is derived in \S\ref{sec:gf} and its applications to the
harmonic and Kepler potentials are described in \S\S\ref{sec:sho} and
\ref{sec:kep}.

\section{A distribution-free estimator based on generating functions}

\label{sec:gf}

\noindent
For simplicity the derivation in this section is for systems with a
single degree of freedom and a single mass parameter. More general
derivations for several degrees of freedom and mass parameters are
given in the Appendix.

We have a set of $N$ particles with positions and velocities
$\{x_n,v_n\}$, $n=1,\ldots,N$.  We choose a trial mass parameter
$\mu_t$ that is (hopefully) close to the true but unknown mass parameter $\mu$,
with $\Delta\mu\equiv \mu-\mu_t$. We compute the actions and angles of
the particles in the potential corresponding to the trial mass
parameter, and call these $\{j_{t,n},\theta_{t,n}\}$. We now seek a function
$P(j_t,\theta_t,\mu_t)$ that yields an estimator $\hat\mu$ of the true
mass $\mu$ through the following formula:
\begin{equation}
\hat\mu=\mu_t +\frac{1}{N}\sum_{n=1}^N
P(j_{t,n},\theta_{t,n},\mu_t).
\label{eq:hatdef}
\end{equation}
We require that the estimator is unbiased to $\mbox{O}(\Delta \mu)$,
by which we mean the following. Suppose that the $N$ particles have a
common action $j$ and a uniform probability distribution in the angle
$\theta$, both variables being defined in the potential corresponding
to the true mass parameter $\mu$. Denoting the average over this
distribution of angles by $\langle\cdot\rangle_\theta$, we require that
\begin{equation}
\langle P(j_t,\theta_t,\mu)\rangle_\theta=\mu-\mu_t +
\mbox{O}(\Delta\mu)^2.
\label{eq:bias}
\end{equation}

The transformation between $(x,v)$ and $(j,\theta)$ or
$(j_t,\theta_t)$ is canonical so the transformation from $(j,\theta)$ to
$(j_t,\theta_t)$ is canonical as well. Therefore it can be described
by a mixed-variable generating function $S(j,\theta_t,\mu_t,\Delta\mu)=j\theta_t +
s(j,\theta_t,\mu_t,\Delta\mu)$ with $s(j,\theta_t,\mu_t,\Delta\mu)=\mbox{O}(\Delta\mu)$ and 
\begin{align}
j_t&=\p_2S(j,\theta_t,\mu_t,\Delta\mu)=j +
\p_2 s(j,\theta_t,\mu_t,\Delta\mu),\nonumber \\
\theta&=\p_1S(j,\theta_t,\mu_t,\Delta\mu)=\theta_t +
\p_1s(j,\theta_t,\mu_t,\Delta\mu).
\label{eq:genfunc}
\end{align}
Here and throughout we use the notation $\p_n f$ to denote the partial
derivative of $f$ with respect to its $n$th argument. The action-angle
variables for mass parameter $\mu_t$ can then be written in terms of
those for mass parameter $\mu$ as
\begin{align}
j_t&=j+\p_2 s(j,\theta,\mu_t,\Delta\mu)+\mbox{O}(\Delta\mu)^2,
\nonumber \\
\theta_t&=\theta-\p_1
s(j,\theta,\mu_t,\Delta\mu)+\mbox{O}(\Delta\mu)^2.
\label{eq:sss}
\end{align}

We can now write
\begin{equation}
P(j_t,\theta_t,\mu_t) = P + (\p_1 P)(\p_2 s) - (\p_2 P)(\p_1 s)
+\mbox{O}(\Delta\mu)^2;
\label{eq:pexp}
\end{equation}
in this formula the arguments of all functions on the right are
$(j,\theta,\mu_t)$ or $(j,\theta,\mu_t,\Delta\mu)$. The condition (\ref{eq:bias}) now reads
\begin{equation}
\langle P \rangle_\theta + \langle (\p_1 P)(\p_2 s) -  (\p_2 P)(\p_1 s)
\rangle_\theta = \Delta\mu +\mbox{O}(\Delta\mu)^2;
\label{eq:ppp}
\end{equation}
To make the dependence on $\Delta\mu$ on the left side explicit, we
expand the generating function $s(j,\theta_t,\mu_t,\Delta\mu)$ as a
power series in $\Delta\mu$,
\begin{equation}
s(j,\theta_t,\mu_t,\Delta\mu)=\Delta\mu\,
s_1(j,\theta_t,\mu_t)+\mbox{O}(\Delta\mu)^2.
\label{eq:genfuna}
\end{equation}
Then equation (\ref{eq:ppp}) becomes 
\begin{equation}
\langle P \rangle_\theta + \Delta\mu\langle (\p_1 P)(\p_2 s_1) - (\p_2 P)(\p_1 s_1)
\rangle_\theta = \Delta\mu +\mbox{O}(\Delta\mu)^2;
\label{eq:pppqqq}
\end{equation}
here the arguments of all functions are $(j,\theta,\mu_t)$. Since $s_1$ is periodic in
the angle variable -- this follows from
equation (\ref{eq:sder}) below -- we may integrate the last term on the
left side by parts. Then (\ref{eq:pppqqq}) can be rewritten as
\begin{equation}
\langle P\rangle_\theta=0, \qquad \p_1\langle P (\p
_2 s_1)\rangle_\theta=1
\label{eq:ttt}
\end{equation}
or
\begin{equation}
\langle P\rangle_\theta=0, \qquad \langle P (\p
_2 s_1)\rangle_\theta=j-j_\star
\label{eq:tttt}
\end{equation}
where $j_\star$ is an integration constant and the arguments of $P$ and
$s_1$ are $(j,\theta,\mu_t)$.  These requirements can be satisfied if
we choose
 \begin{equation}
P(j,\theta,\mu_t)= P_\star(j,\theta,\mu_t)\equiv \frac{(j-j_\star)\,\p_2 s_1(j,\theta,\mu_t)}{\langle [\p_2 s_1(j,\theta,\mu_t)]^2\rangle_\theta};
\label{eq:pdef}
\end{equation}
we call these generating-function estimators. Other functions
can satisfy the constraints (\ref{eq:tttt}) but $P_\star$ is optimal
in a sense that we now describe.

If the trial mass is fixed and equal to the true mass, then the
variance in the estimator $\hat\mu$ is
\begin{align}
\sigma^2& \equiv \big\langle(\hat\mu-\langle\hat\mu\rangle_\theta)^2\big\rangle_\theta=\langle\hat\mu^2\rangle_\theta
-\langle\hat\mu\rangle_\theta^2 \nonumber \\
&=\frac {1}{N^2}\sum_{n=1}^N\left[\langle
  P^2(j_{n},\theta,\mu)\rangle_\theta -\langle
  P(j_{n},\theta,\mu)\rangle_\theta^2\right].
\label{eq:vardef1}
\end{align}
Now $\langle P\rangle_\theta=0$ by (\ref{eq:ttt}), so 
\begin{equation}
\sigma^2=\frac {1}{N^2}\sum_{n=1}^N\langle
  P^2(j_{n},\theta,\mu)\rangle_\theta.
\label{eq:vardef2}
\end{equation}
Now write $P=P_\star+\Delta P$. Then $\langle P^2\rangle_\theta=\langle
P_\star^2\rangle_\theta + 2(j-j_\star)\langle (\Delta P)(\p_2
s_1)\rangle_\theta/\langle(\p_2s_1)^2\rangle_\theta +\langle\Delta P^2\rangle_\theta$. The
second of equations (\ref{eq:tttt}) requires $\langle (\Delta P)(\p_2
s_1)\rangle_\theta=0$. Therefore $\langle P^2\rangle_\theta=\langle
P_\star^2\rangle_\theta +\langle\Delta P^2\rangle_\theta$; this means that $P_\star$
has the smallest variance among all estimators with a given
value of the integration constant $j_\star$. For this reason we adopt the
estimator $P_\star$ in preference to any other estimators that satisfy
equations (\ref{eq:tttt}). The variance is then
\begin{equation}
\sigma^2=\frac {1}{N^2}\sum_{n=1}^N\frac{(j_n-j_\star)^2}{\langle
[\p_2 s_1(j_n,\theta,\mu)]^2\rangle_\theta}.
\label{eq:vardef2}
\end{equation}

The estimator is distribution-free in the sense that we have made no
assumptions about the \df\ $F(\bfj)$ in deriving it. We have no
general proof that the estimator is always consistent, but we show
below that the GF0 estimator is consistent for the harmonic and
Kepler potentials. 

We are free to choose the constant $j_\star$. The simplest approach
is to set $j_\star=0$\footnote{In principle, this involves no loss of
  generality because there are action-angle variables $(j',\theta')$
  in which $j'=j-j_\star$ and $\theta'=\theta$.}, and an estimator
based on this choice will be called a GF0 estimator. A potentially more powerful approach
is to choose $j_\star$ to minimize the variance $\sigma^2$. From
equations (\ref{eq:pdef}) and (\ref{eq:vardef2}) this condition
implies
\begin{equation}
j_\star=j_{\rm min}\equiv \frac{\sum_n j_n/\langle [\p_2
  s_1(j,\theta,\mu_t)]^2\rangle_\theta}{\sum_n 1/\langle [\p_2
  s_1(j,\theta,\mu_t)]^2\rangle_\theta}.
\label{eq:jstar-def}
\end{equation}
An estimator based on this choice for $j_\star$ will be called a GF1
estimator.  Of course, initially we do not know the true actions and
angles so we must use an iterative procedure to determine
$j_\mathrm{min}$. An approach that works well in our experiments is to
(i) determine an estimate $\hat\mu$ for the mass parameter using
$j_\star=0$; (ii) use this estimate to compute actions and angles, and 
insert these in equation (\ref{eq:jstar-def}) to determine
$j_{\rm min}$; (iii) use $j_\star=j_{\rm min}$ to derive an improved
estimate $\hat\mu$.

The formula (\ref{eq:hatdef}) that determines the
estimated mass $\hat\mu$ in terms of the trial mass $\mu_t$ can be
iterated to convergence, which occurs when $\hat\mu=\mu_t$ or 
\begin{equation}
\sum_{n=1}^NP(j_{\hat\mu,n},\theta_{\hat\mu,n},\hat\mu)=0.
\label{eq:conv}
\end{equation}
This is a non-linear equation for $\hat\mu$ and it may have more than
one solution. We have no general procedure for choosing the correct
solution but show how to do so for the harmonic and Kepler
potentials in \S\S \ref{sec:sho} and \ref{sec:kep} respectively. Note 
that:

\begin{enumerate}

\item The generating-function estimators are unbiased only for a fixed
  value of the trial mass, so some bias is introduced when the
  estimator (\ref{eq:conv}) is used. In our experiments this bias is
  always small compared to the standard deviation of the estimator
  (see Figures \ref{fig:one} and \ref{fig:three}).

\item The formula (\ref{eq:vardef2}) for the variance is valid for a
  fixed value of the trial mass, and if we iterate to convergence
  using (\ref{eq:conv}) the variance in $\hat\mu$ will generally be
  larger.

\end{enumerate}

\subsection{The generating function}

\noindent
To implement this method we need the first-order generating function
$s_1(j,\theta_t)$ for the canonical transformation between
$(j,\theta)$ and $(j_t,\theta_t)$ (eq.\ \ref{eq:genfuna}). We assume
that the Hamiltonian has the usual form $H(x,v|\mu)=\half
v^2+\Phi(x|\mu)$. Then in the trial action-angle variables we have
\begin{equation}
H(j_t|\mu_t) = H(j|\mu) +\Phi(x|\mu_t)-\Phi(x|\mu). 
\end{equation}
Using equations (\ref{eq:sss})  and working to first order in
$\Delta\mu=\mu-\mu_t$,
\begin{equation}
\frac{\p H}{\p\mu} -\frac{\p H}{\p j}\frac{\p
  s_1}{\p\theta}=\frac{\p\Phi}{\p\mu}.
\label{eq:kkk}
\end{equation}
Any function $g(j,\theta,\mu)$ can be split into angle-averaged and oscillating parts,
\begin{equation}
\langle g\rangle_\theta = \frac{1}{2\upi}\int_0^{2\upi}
d\theta\, g(j,\theta,\mu), \qquad \{ g \}_\theta\equiv g-\langle g\rangle_\theta.
\end{equation}
Equation (\ref{eq:kkk}) can be re-arranged as
\begin{equation}
\frac{\p H}{\p\mu} -\left\langle \frac{\p\Phi}{\p\mu} \right\rangle_\theta=
\left\{\frac{\p\Phi}{\p\mu}\right\}_\theta+\frac{\p H}{\p
  j}\frac{\p s_1}{\p\theta}.
\end{equation} 
The terms on the left side are independent of angle, and the terms on
the right average to zero. Therefore the left and right
sides must both be zero, and we have 
\begin{equation}
\frac{\p H}{\p\mu} =\left\langle \frac{\p\Phi}{\p\mu} \right\rangle_\theta,
\quad s_1(j,\theta,\mu)=-\left(\frac{\p H}{\p
    j}\right)^{-1}\int^\theta \ud\theta \,\left\{\frac{\p\Phi}{\p\mu}\right\}_\theta. 
\label{eq:sder}
\end{equation}
The second equation implies that $s_1(j,\theta)$ is a periodic
function of $\theta$, a result we used in deriving (\ref{eq:pdef}).

\section{The harmonic potential}

\label{sec:sho}

\noindent
We now apply the generating-function estimator to particles orbiting in the one-dimensional harmonic potential,
\begin{equation}
\Phi(x|\,\omega)=\half\omega^2 x^2.
\label{eq:phisho}
\end{equation}
Here the mass parameter is $\omega$, the oscillator frequency.  The
Hamiltonian is
\begin{equation}
H(x,v|\,\omega)=\half v^2 + \half\omega^2 x^2.
\label{eq:hsho1}
\end{equation}
The solution of the equations of motion is
\begin{equation}
x=\sqrt{\frac{2j}{\omega}}\cos\theta , \quad
  v=\sqrt{2j\omega}\sin\theta; 
\label{eq:t2}
\end{equation}
here $(j,\theta)$ are the action-angle variables for
frequency $\omega$ and in these variables the Hamiltonian is
\begin{equation}
H(j|\omega)=\omega j. 
\label{eq:hsho}
\end{equation}

Suppose that the true value of the frequency is $\omega$ and
$\omega_t$ is a trial approximation to $\omega$. We denote the action
and angle corresponding to $\omega_t$ by $j_t$ and $\theta_t$. The
relation between the action-angle pairs $(j,\theta)$ and
$(j_t,\theta_t)$ is found by solving
\begin{equation}
x=\sqrt{\frac{2j}{\omega}}\cos\theta=\sqrt{\frac{2j_t}{\omega_t}}\cos\theta_t,
\ \ 
v=\sqrt{2j\omega}\sin\theta=\sqrt{2j_t\omega_t}\sin\theta_t,
\label{eq:actang}
\end{equation}
which yields
\begin{equation}
\theta=\tan^{-1}\frac{\omega_t}{\omega}\tan\theta_t, \quad
j=j_t\left(\frac{\omega}{\omega_t}\cos^2\theta_t+\frac{\omega_t}{\omega}\sin^2\theta_t\right). 
\label{eq:sho}
\end{equation}
The generating function (\ref{eq:genfuna}) for this transformation is
\begin{equation}
s(j,\theta_t,\omega_t,\Delta\omega)=j\tan^{-1}\frac{\omega_t}{\omega_t+\Delta\omega}\tan\theta_t
-j\theta_t,
\end{equation}
where $\Delta\omega=\omega-\omega_t$. Expanding as a power series in
$\Delta\omega$, we have
$s(j,\theta_t,\omega_t,\Delta\omega)=\Delta\omega
s_1(j,\theta_t,\omega_t)+\mbox{O}(\Delta\omega)^2$ where
\begin{equation}
s_1(j,\theta,\omega)=-\frac{j}{\omega}\sin\theta\cos\theta.
\label{eq:s1def}
\end{equation}

This result can also be derived from equation (\ref{eq:sder}). From
equation (\ref{eq:phisho}) $\p \Phi(x|\omega)/\p\omega=\omega x^2$ and
from equation (\ref{eq:t2})
\begin{equation}
\left\langle\frac{\p\Phi}{\p\omega}\right\rangle_\theta=\langle
2j\cos^2\theta\rangle_\theta = j, \quad
\left\{\frac{\p\Phi}{\p\omega}\right\}_\theta=j
(2\cos^2\theta-1).
\end{equation}
Equation (\ref{eq:hsho}) yields $\p H/\p j=\omega$ so
\begin{equation}
s_1(j,\theta,\omega)= -\frac{j}{\omega}\int^\theta \ud\theta\,(2\cos^2\theta-1)=
  -\frac{j}{\omega}\sin\theta\cos\theta,
\end{equation}
consistent with (\ref{eq:s1def}).

To derive the generating-function estimator of the frequency we need the result
\begin{equation}
\langle [\p_2 s_1(j,\theta,\omega)]^2\rangle_\theta=\frac{j^2}{2\omega^2}.
\end{equation} 
From equation (\ref{eq:pdef}) we then have 
\begin{equation}
P_\star(j,\theta,\omega)=-2\omega\,\frac{j-j_\star}{j}\cos 2\theta.
\end{equation}
The estimator (\ref{eq:hatdef}) is then 
\begin{equation}
\hat\omega=\omega_t\bigg[1-\frac{2}{N}\sum_{n=1}^N\left(1-\frac{j_\star}{j_{t,n}}\right)\cos 2\theta_{t,n}\bigg].
\label{eq:maxl-sho}
\end{equation}
We may choose either $j_\star=0$ for simplicity, which yields the GF0
estimator for the harmonic potential, or (eq.\ \ref{eq:jstar-def})
\begin{equation}
j_\star=j_{\rm min}=\frac{\sum_n j_{t,n}^{-1}}{\sum_n j_{t,n}^{-2}}
\label{eq:jmin-h}
\end{equation}
which yields the minimum-variance or GF1 estimator. 

The variance in $\hat\omega$ at a fixed value of the trial frequency
can be determined from equation (\ref{eq:vardef2}): If the trial
frequency $\omega_t$ is equal to the true frequency $\omega$ then
\begin{equation}
\sigma^2=\frac{2\omega^2}{N^2}\sum_{n=1}^N\left(1-\frac{j_\star}{j_n}\right)^2.
\label{eq:sho-sig}
\end{equation}
When $j_\star=0$, $\sigma^2=2\omega^2/N$. Since $\sigma^2$ is finite
and declines as $N^{-1}$ the GF0 estimator is consistent, and by the
central-limit theorem the distribution of estimates $\hat\omega$ is
Gaussian for large $N$. 

For the minimum-variance estimator GF1, $j_\star$ is given by equation
(\ref{eq:jmin-h}) and the variance is
\begin{equation}
\sigma^2=\frac{2\omega^2}{N}\left(1-\frac{\langle
    j^{-1}\rangle^2}{\langle j^{-2}\rangle}\right).
\label{eq:vargf1}
\end{equation} 

We now iterate this procedure, replacing the trial frequency $\omega_t$
by $\hat\omega$, re-evaluating the angles $\theta_n$, and
repeating the process. We have converged when
$\hat\omega=\omega_t$, which requires
\begin{equation}
\sum_{n=1}^N \left(1-\frac{j_\star}{j_{\hat\omega,n}}\right)\cos
2\theta_{\hat\omega,n}=0
\end{equation}
or
\begin{equation} 
\sum_{n=1}^N\frac{(\hat\omega^2x_n^2+v_n^2-2\hat\omega j_\star)(\hat\omega^2x_n^2-v_n^2)}{(\hat\omega^2x_n^2+v_n^2)^2}=0.
\label{eq:msho}
\end{equation}

When $j_\star=0$ the estimator (\ref{eq:msho}) simplifies to
\begin{equation}
\sum_{n=1}^N\frac{\hat\omega^2x_n^2-v_n^2}{\hat\omega^2x_n^2+v_n^2}=0.
\label{eq:mshoo}
\end{equation}
The left side approaches $-N$ as $\hat\omega\to 0$ and approaches $+N$
as $\hat\omega\to\infty$ and is monotonic in $\hat\omega$. Therefore
(\ref{eq:mshoo}) has one and only one solution for $\hat\omega$. When $j_\star\not=0$
the asymptotic behavior of the left side is the same; thus there is at
least one solution in the range $0<\hat\omega<\infty$ but there may be
more than one solution. To determine which one to use in practice, we first find
the unique solution $\hat\omega_0$ with $j_\star=0$ from
(\ref{eq:mshoo}), then find all the solutions with the given value of
$j_\star$ from (\ref{eq:msho}) and choose the one that is closest to
$\hat\omega_0$.

For the harmonic potential, but not in general, the
variance in $\hat\omega$ at a fixed value of the trial frequency is the
same as the variance in $\hat\omega$ as determined by iterating to
convergence. 

\subsection{A maximum-likelihood estimator} 

\label{sec:maxl}

\noindent
The harmonic potential is one of the few for which a
calculation of the likelihood can be carried out analytically without
assumptions about the \df, and this result can be compared with the GF
estimators derived above.

Let  $F(x,v)=F(j)$ be the \df, defined in \S\ref{sec:intro} and normalized so $\int
F(j)djd\theta=2\upi\int F(j)dj\- =1$.  Then the probability
that a particle lies in the small phase-space volume $dxdv$ is
$F(j)dxdv\-= F(\half v^2/\omega + \half \omega x^2)dxdv$. Now let $z\equiv
shov/x$ and ask for the probability density of $z$:
\begin{align}
p(z)&=\int dx dv F(\half v^2/\omega + \half \omega
x^2)\delta(z-v/x) \nonumber \\ 
&=\int dx\, xF(\half x^2z^2/\omega + \half \omega
x^2)\nonumber \\
&=\frac{2\omega}{z^2+\omega^2}\int F(j)
dj=\frac{\omega}{\upi(z^2+\omega^2)}.
\label{eq:maxlz}
\end{align}
Given a data set $\{ z_n\}$,  the log-likelihood is then \citep[cf.][]{mag14}
\begin{equation}
\log L=\mbox{const} + N\log\omega -
\sum_{n=1}^N\log({z_n}^2+\omega^2).
:\label{eq:sho-l}
\end{equation}
The maximum-likelihood estimate of the frequency, $\hat\omega$, is
given by the solution of $\p\log L/\p \omega=0$, which is
\begin{equation}
\sum_{n=1}^N
\frac{\hat\omega^2-{z_n}^2}{\hat\omega^2+{z_n}^2}=0
\quad\mbox{or} \quad \sum_{n=1}^N
\frac{\hat\omega^2x_n^2-v_n^2}{\hat\omega^2x_n^2+v_n^2}=0.
\label{eq:maxl-exact}
\end{equation}
The variance is given by
\begin{equation}
\frac{1}{\sigma^2}=-\left\langle \frac{\p^2\log
    L}{\p\omega^2}\right\rangle_z=-\frac{N}{2\omega^2};
\label{eq:var-maxl}
\end{equation}
here $\langle \cdot\rangle_z$ denotes the average over the probability
density (\ref{eq:maxlz}). 

Despite its simplicity, this method is of limited usefulness for
two reasons. First, the derivation works only for a limited set of
potentials of the form $\Phi(x)=c|x|^\alpha$, and cannot be generalized
to arbitrary one-dimensional systems or to three-dimensional
systems such as the Kepler potential. Second, by fitting only the
distribution of $z=v/x$ the method ignores the distribution of
actions, which also helps to constrain the frequency $\omega$ -- for
example, at the correct frequency there should be no correlation
between the actions and angles \citep{bl04}. 

The estimator (\ref{eq:maxl-exact}) and its variance
(\ref{eq:var-maxl}) are identical to those of the GF0 method,
equations (\ref{eq:mshoo}) and (\ref{eq:sho-sig}) with $j_\star=0$.
Thus for the harmonic potential, the GF0 estimator is equivalent to
maximizing the likelihood of the data set $\{ z_n\}$.

\subsection{Numerical tests}

\label{sec:sho-num}

\noindent
To explore the performance of various estimators for the frequency
$\omega$ of a harmonic potential, we generate 5000 realizations
of the positions and velocities $\{x_n,v_n\}$ of $N$ particles
orbiting in the potential (\ref{eq:phisho}) with frequency $\omega=1$. The
particles are distributed uniformly random in phase, and randomly in
the amplitude $A\equiv \sqrt{2j/\omega}$ (eq.\ \ref{eq:t2}) with
probability density
\begin{equation}
dp\propto A^{-\gamma} \,d\log A , \qquad A_{\rm min} < A < A_{\rm max}
\label{eq:dist}
\end{equation}
and zero otherwise. 

We compare four estimators: (i) the virial theorem (eq.\
\ref{eq:virial-h}); (ii) the maximum-likelihood estimator
(\ref{eq:maxl-exact}), which for the harmonic potential is equivalent
to the GF0 estimator with $j_\star=0$ (eq.\ \ref{eq:mshoo}); (iii)
orbital roulette, using the Anderson--Darling test for uniformity of
the orbital phases \citep{bl04}\footnote{As we observed after equation
  (\ref{eq:uniform}), the version of orbital roulette based on the
  mean phase fails for the harmonic potential.}; (iv) the
minimum-variance GF1 estimator (eqs.\ \ref{eq:jmin-h} and
\ref{eq:msho}).

\begin{figure}
\vspace{-0.0cm}
\centering
\includegraphics[width=\columnwidth]{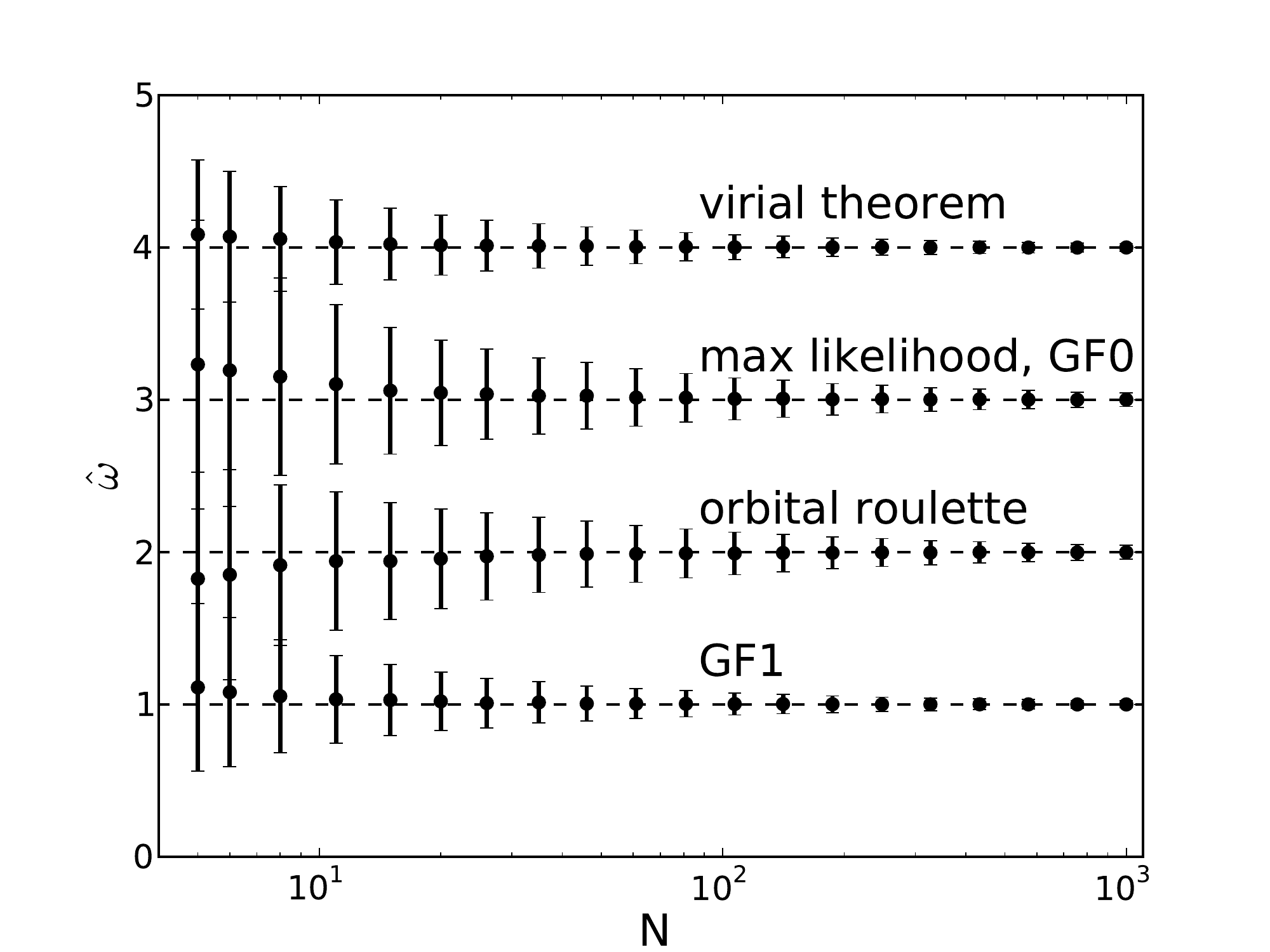}
\caption {\small Performance of four estimators for $N$ particles
  orbiting in a harmonic potential. From top to bottom, the
  estimators are the virial theorem (eq.\ \ref{eq:virial-h}), the
  maximum-likelihood estimator (\ref{eq:maxl-exact}), orbital
  roulette, and the minimum-variance GF1 estimator (eqs.\
  \ref{eq:jmin-h} and \ref{eq:msho}). The mean estimate for the
  frequency $\omega$ and the standard deviation are determined for
  5000 realizations and shown by the filled circles and error
  bars. The true frequency is $\omega=1$, and the distribution of
  amplitudes is given by equation (\ref{eq:dist}) with $\gamma=0$ and
  $A_{\rm max}/A_{\rm min}=3$. To minimize confusion, the results for
  the different estimators have been displaced by integer offsets on
  the vertical axis.}
\label{fig:one}
\end{figure}

The mean and standard deviation of the estimators $\hat\omega$ are
shown in Figure \ref{fig:one} over the range $N=5$--1000, for a
distribution of amplitudes having $\gamma=0$ and $A_{\rm max}/A_{\rm
  min}=3$. All four estimators exhibit modest bias for small $N$, but
are consistent in the sense that the bias and standard deviation
approach zero as the sample size $N\to\infty$. At
$N=1000$, the standard deviations in $\hat\omega$, ordered
by increasing magnitude, are: GF1, 0.023; then virial theorem, 0.026,
then GF0, maximum likelihood, and orbital roulette tied at 0.045 (for
all methods except roulette, these results can be derived
analytically; see equations \ref{eq:sho-var}, \ref{eq:sho-sig}, \ref{eq:vargf1},  
and \ref{eq:var-maxl}).

\begin{figure}
\vspace{-0.0cm}
\centering
\includegraphics[width=\columnwidth]{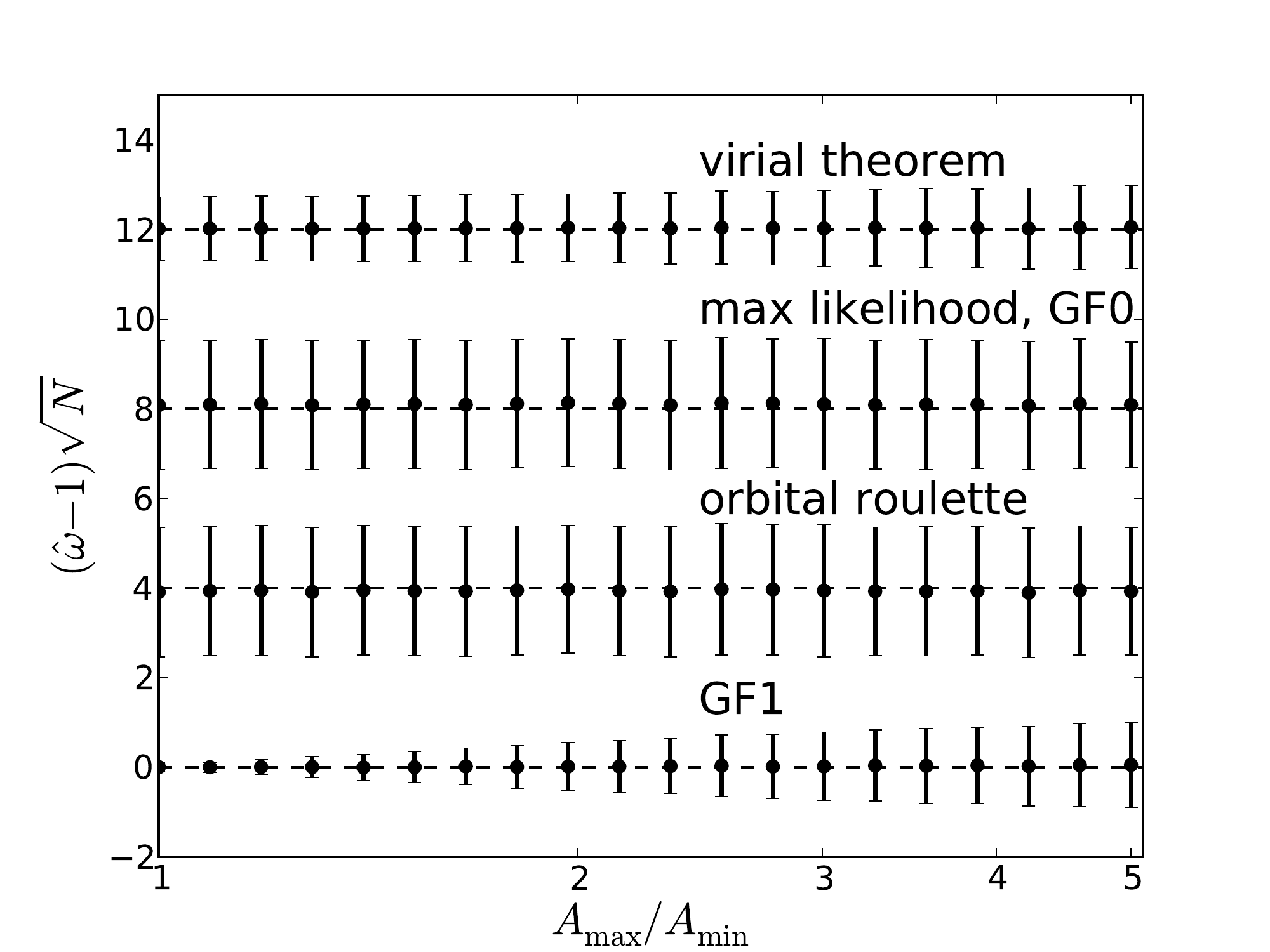}
\caption {\small Performance of the same four estimators as in Figure
  \ref{fig:one} for $N=100$ particles orbiting in
  a harmonic potential with $\omega=1$. The distribution of
  amplitudes is given by equation (\ref{eq:dist}) with $\gamma=0$ and
  varying $A_{\rm max}/A_{\rm min}$ (horizontal axis). To minimize confusion, the
  results for the different estimators have been displaced by 
  integer offsets on the vertical axis. We plot
  $(\hat\omega-1)\surd{N}$, which is  independent of $N$ for $N\gg1$.}
\label{fig:two}
\end{figure}

Figure \ref{fig:two} shows the performance of the estimators for fixed
sample size $N=100$, $\gamma=0$, and a range of values for
$A_\mathrm{max}/A_\mathrm{min}$. Rather than plotting $\hat\omega$ as
in Figure \ref{fig:one}, we plot $(\hat\omega-1)\surd{N}$, which is
independent of sample size. The behavior of the maximum-likelihood and
orbital-roulette estimators is independent of the distribution of
amplitudes $A$
since they depend on the data only
through the ratio $z=v/x$ (eq.\ \ref{eq:sho-l}) or $\theta$ (eq.\
\ref{eq:actang}) whose distributions are independent of the amplitude
(this independence holds only for the harmonic potential, not for
most others). In this figure, the standard deviation of the
generating-function estimator GF1 is always the smallest of the four
estimators. It is remarkable that the estimator GF1 performs so much
better than the others when $A_\mathrm{max}/A_{\rm min}\sim
1$. 

\section{The Kepler potential}

\label{sec:kep}

\noindent
The Hamiltonian for the Kepler potential is
\begin{equation}
H(\bfx,\bfv|\mu)=\half v^2+\Phi(r|\mu)=\half v^2 -\frac{\mu}{r}
\end{equation}
where $r=|\bfx|$, $\mu=GM$, $G$ is the gravitational constant, and we
loosely refer to $\mu$ as the mass of the central body.

An orbit is described by its semimajor axis $a$, eccentricity
$e$, inclination $i$ relative to some $x$-$y$ reference plane, longitude of
node $\Omega$, argument of periapsis $\omega$, and mean anomaly
$\ell$. The specific angular momentum is
$L=|\bfx\times\bfv|=[\mu a(1-e^2)]^{1/2}$ and the $z$-component of the
specific angular momentum is $L_z=L\cos i$. A suitable set of
action-angle variables is
\begin{align}
\bfj&=(j_1,j_2,j_3)=\sqrt{\mu a}\Big(1-\sqrt{1-e^2},\sqrt{1-e^2},\sqrt{1-e^2}\cos
i\Big) \nonumber \\
&=\Big(\sqrt{\mu a}\Big[1-\sqrt{1-e^2}\Big],L,L_z\Big), \nonumber \\
\bftheta&=(\theta_1,\theta_2,\theta_3)=(\ell,\ell+\omega,\Omega);
\label{eq:aa}
\end{align}
$j_1$ is called the radial action since it vanishes for circular
orbits. In these variables the Hamiltonian is $H=-\half\mu^2/(j_1+j_2)^2$. 
We shall also use the eccentric anomaly $u$, given by Kepler's
equation $\ell=u-e\sin u$.

Since the orientation of the orbit does not help to constrain the mass
$\mu$, we can ignore the angles $\theta_2$ and $\theta_3$ and the action
$L_z$. Moreover the angular momentum $L=|\bfx\times\bfv|$ is
independent of the mass $\mu$ for fixed position and velocity
$(\bfx,\bfv)$. Therefore the mass-dependent dynamics has only one
degree of freedom and is described by the action $j_1$ and its
conjugate angle $\theta_1$. For brevity we now drop the
subscript so $j_1\to j$ and $\theta_1\to\theta=\ell$. 

We have $\p\Phi/\p\mu=-1/r$, $r=a(1-e\cos u)$, and $d\ell=(1-e\cos u)du$
so
\begin{equation}
\left\langle\frac{\p\Phi}{\p \mu}\right\rangle_\theta=-\int_0^{2\upi}
    \frac{d\ell}{2\upi}\frac{1}{a(1-e\cos u)}=-\frac{1}{a},
\end{equation}
and 
\begin{equation}
\left\{\frac{\p\Phi}{\p
    \mu}\right\}_\theta=-\frac{1}{r}+\frac{1}{a}=-\frac{e\cos u}{a(1-e\cos u)}.
\end{equation}
Then equation (\ref{eq:sder}) yields the generating function
\begin{align}
s_1(j,\theta,\mu)=\frac{(j+L)^3}{\mu^2}\int^u du(1-e\cos
u)\left\{\frac{\p\Phi}{\p\mu}\right\}_\theta=\sqrt{\frac{a}{\mu}}e\sin u,
\end{align}
and the results 
\begin{align}
\p_2 s_1(j,\theta,\mu)&=\sqrt{\frac{a}{\mu}}\frac{e\cos
  u}{1-e\cos u}, \nonumber \\
\langle[\p_2 s_1(j,\theta,\mu)]^2\rangle_\theta&=\frac{a}{\mu}\left[(1-e^2)^{-1/2}-1\right].
\end{align}
In these formulae $u$, $e$, and $a$ should be regarded as functions of
the action $j=(\mu a)^{1/2}[1-(1-e^2)^{1/2}]$, the angle
$\theta=\ell=u-e\sin u$, and the angular momentum
$L=[\mu a(1-e^2)]^{1/2}$. From equations (\ref{eq:hatdef}) and
(\ref{eq:pdef}), the generating-function estimator of the mass is
given by
\begin{equation}
\frac{\hat\mu}{\mu_t}=1+\frac{1}{N}\sum_{n=1}^N\left(1-\frac{j_\star}{j_{t,n}}\right)\frac{e_{t,n}(1-e_{t,n}^2)^{1/2}\cos
  u_{t,n}}{1-e_{t,n}\cos  u_{t,n}},
\label{eq:maxl-kep}
\end{equation}
The estimator can be re-written in terms of observable quantities, the
radius $r$, speed $v$, and radial and tangential velocities
$v_r$ and $v_\perp=(v^2-v_r^2)^{1/2}$. In making this conversion two useful
identities are 
\begin{equation}
v^2r=\mu(1+e\cos u), \qquad
v_r^2r=\mu\frac{e^2\sin^2u}{1-e\cos u}.
\label{eq:vr}
\end{equation}
We find
\begin{align}
\frac{\hat\mu}{\mu_t}=1+\frac{1}{N}\sum_{n=1}^N\left(1-\frac{j_\star}{j_{t,n}}\right) 
\frac{(v_n^2r_n-\mu_t)v_{\perp,n}r_n^{1/2}}{\mu_t(2\mu_t-v_n^2r_n)^{1/2}}.
\label{eq:maxl-kep1}
\end{align}
where
\begin{equation}
j_{t,n}=\left(\frac{r_n}{2\mu_t-v_n^2r_n}\right)^{1/2}\big[\mu_t-(2\mu_t -
v_n^2r_n)^{1/2}v_{\perp,n}r_n^{1/2}\big].
\end{equation}
The quantities within the square roots are positive so long as
$2\mu_t\ge v_n^2r_n$, which holds so long as the particle is in a
bound orbit at the trial mass.

The variance of the estimator $\hat\mu$ can be determined from
equation (\ref{eq:vardef1}) or (\ref{eq:maxl-kep}). If the trial mass
equals the true mass, the variance at a fixed value of the trial mass
is
\begin{align}
    \sigma^2&=\big\langle(\hat\mu-\langle\hat\mu\rangle_\theta)^2\big\rangle_\theta\nonumber
    \\
&=\frac{\mu^2}{N^2}\sum_{n=1}^N \left(1-\frac{j_\star}{j_{n}}\right)^2
(1-e^2_n)^{1/2}\big[1-(1-e_n^2)^{1/2}\big]. 
\label{eq:vardef-kep}
\end{align}
When $j_\star=0$, the variance is bounded above by $\mu^2/(4N)$, which
implies that the estimator is consistent, and by the central-limit
theorem the distribution of estimates $\hat\mu$ is Gaussian for large
$N$.

The variance is minimized when we choose
\begin{align}
j_\star&=j_\mathrm{min}=\mu^{1/2}\frac{\sum_n (1-e_n^2)^{1/2}a_n^{-1/2}}{\sum_n (1-e_n^2)^{1/2}a_n^{-1}\big[1-(1-e_n^2)^{1/2}\big]^{-1}}
\label{eq:jmin-m}\\
&=\frac{\sum_n (2\mu-v_n^2r_n)v_{\perp,n}}{\sum_n (2\mu-v_n^2r_n)^{3/2}v_{\perp,n}
  \big[\mu r_n^{1/2}-(2\mu-v_n^2r_n)^{1/2}v_{\perp,n}r_n\big]^{-1}}.
\nonumber
\end{align}

The estimate has converged when $\hat\mu=\mu_t$, which occurs when
\begin{equation}
\sum_{n=1}^N\left(1-\frac{j_\star}{j_{\hat\mu,n}}\right)
\frac{(v_n^2r_n-\hat\mu)v_{\perp,n}r_n^{1/2}}{(2\hat\mu-v_n^2r_n)^{1/2}}=0.
\label{eq:maxl-kep2}
\end{equation}

When $j_\star=0$  the estimator (\ref{eq:maxl-kep2}) simplifies to
\begin{equation}
\sum_{n=1}^N \frac{(v_n^2r_n-\hat\mu)v_{\perp,n}r_n^{1/2}}{(2\hat\mu-v_n^2r_n)^{1/2}}=0,
\label{eq:mkep0}
\end{equation}
and we denote the estimator based on this formula as GF0. We denote
the minimum-variance estimator based on equations (\ref{eq:maxl-kep2}) and
(\ref{eq:jmin-m}) as GF1.  Note that the generating-function estimators
are unbiased, and their variance is given by (\ref{eq:vardef-kep}), only
for a fixed value of the trial mass. When iterated to convergence
there will generally be a small non-zero bias, and the variance will
be larger.

The smallest allowed mass is $\hat\mu_{\rm min}=\half v^2_mr_m$ where
$m$ is the index corresponding to the smallest of $\{ v_n^2r_n\}$. As
$\hat\mu\to \hat\mu_{\rm min}$ from above, the left side of
(\ref{eq:mkep0}) approaches
$2^{-3/2}v_m^2v_{\perp,m}r_m^{3/2}(\hat\mu-\hat\mu_{\rm min})^{-1/2}\to
+\infty$. As $\hat\mu\to\infty$, the left side approaches
$-2^{-1/2}{\hat\mu}^{1/2}\sum_n v_{\perp,n}r_n^{1/2}\to-\infty$. Moreover the left
side is a monotonically decreasing function of $\hat\mu$. Therefore
(\ref{eq:mkep0}) has one and only one solution for $\hat\mu$.

When $j_\star\not=0$ there may be more than one solution to
(\ref{eq:maxl-kep2}). To determine which one to use, we first find
the unique solution $\hat\mu_0$ with $j_\star=0$, then find all the
solutions with the chosen value of $j_\star$ and choose the one that
is closest to $\hat\mu_0$.

These results are derived from the action-angle variables
(\ref{eq:aa}). Other sets of action-angle variables yield different
estimators. For example, if we choose Delaunay elements
\begin{align}
\bfj&=\sqrt{\mu a}\Big(1,\sqrt{1-e^2},\sqrt{1-e^2}\cos
i\Big)=(\sqrt{\mu a},L,L_z), \nonumber \\
\bftheta&=(\ell,\omega,\Omega),
\label{eq:aad}
\end{align}
the estimator analogous to (\ref{eq:mkep0}) is
\begin{equation}
\sum_{n=1}^N\frac{(v_n^2r_n-\hat\mu)v_{\perp,n}r_n^{1/2}}{(2\hat\mu-v_n^2r_n)^{1/2}\big[\hat\mu  -v_{\perp,n}r_n^{1/2}(2\hat\mu-v_n^2r_n)^{1/2}\big]}=0.
\label{eq:maxl-kep-alt}
\end{equation}
In general this performs less well than (\ref{eq:mkep0}). For example,
if the estimated mass is correct and the orbits are circular, then
$\hat\mu=\mu$ and $v_n^2r_n=v_{\perp,n}^2r_n=\mu$, so both the numerator and
denominator in each term of (\ref{eq:maxl-kep-alt}) vanish. 

\subsection{Numerical tests}

\noindent
Our tests are similar to those in \S\ref{sec:sho-num} for the
harmonic potential. We generate 5000 realizations of the
positions and velocities $\{\bfx_n,\bfv_n\}$ of $N$ particles orbiting
in a point-mass potential with unit mass, $\mu=1$. The particles are
distributed uniformly random in phase, and randomly in
the semimajor axis $a$ with
probability density
\begin{equation}
dp\propto a^{-\gamma} \,d\log a, \qquad a_{\rm min} < a < a_{\rm max},
\label{eq:dist-a}
\end{equation}
and zero otherwise. We have conducted tests with a variety of
eccentricity distributions.  We compare several estimators:

\begin{figure}
\vspace{-0.0cm}
\centering
\includegraphics[width=\columnwidth]{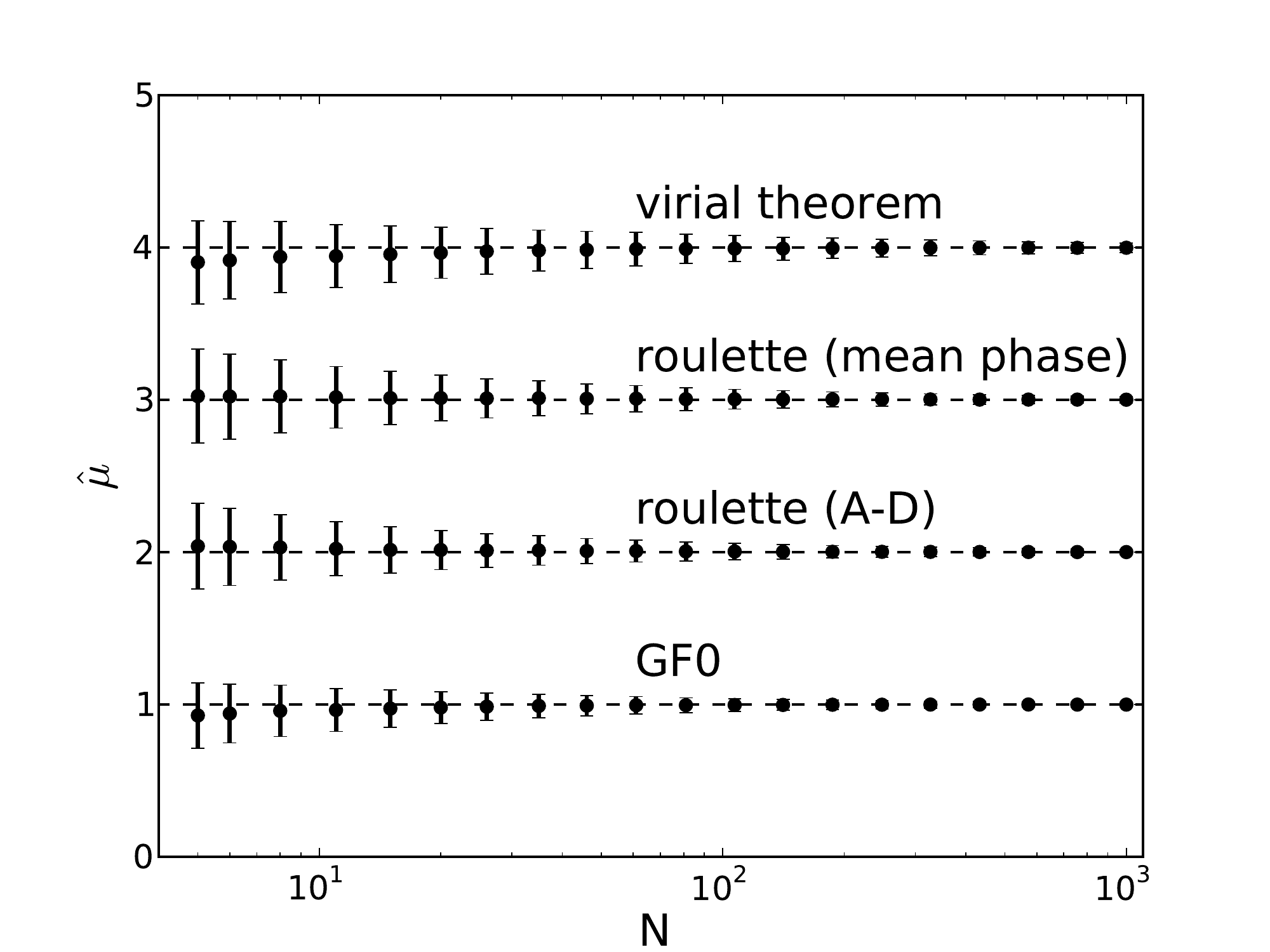}
\caption {\small Performance of estimators for $N$ particles orbiting
  in a Kepler potential with mass $\mu=1$. The squared eccentricities
  are distributed uniformly random between 0 and 1. From top to
  bottom, the estimators are the virial theorem (eq.\
  \ref{eq:virial-m}), orbital roulette using the mean phase, orbital
  roulette using the Anderson--Darling test, and the simple
  generating-function estimator GF0 ($j_\star=0$; eq.\
  \ref{eq:mkep0}). The mean estimate for the mass and the standard
  deviation in $\hat\mu$ are determined for 5000 realizations and
  shown by the filled circles and error bars. The distribution of
  semimajor axes is given by equation (\ref{eq:dist-a}) with
  $\gamma=0$ and $a_{\rm max}/a_{\rm min}=3$. To minimize confusion,
  the results for the different estimators have been displaced by
  integer offsets on the vertical axis.}
\label{fig:three}
\end{figure}

\begin{enumerate}

\item the virial theorem (eq.\ \ref{eq:virial-m}); 

\item orbital roulette, using both the mean-phase test and the
  Anderson--Darling test as described by \cite{bl04}; 

\item the simple generating-function estimator GF0 ($j_\star=0$, eq.\
  \ref{eq:mkep0}) and the minimum-variance estimator GF1
  ($j_\star=j_{\rm min}$, eqs.\ \ref{eq:maxl-kep2} and
  \ref{eq:jmin-m});

\item if the particles have a known eccentricity $e$ then two
  unbiased estimators of the mass are 
\begin{align}
\label{eq:meanx}
\hat\mu &=\frac{1}{N(1-\half e^2)}\sum_{n=1}^N v_n^2r_n, \\
\hat\mu &=\frac{2}{Ne^2}\sum_{n=1}^N v_{r,n}^2r_n,
\label{eq:meany}
\end{align}
These estimators are not useful in practice since the eccentricities
are not known, but they provide a useful benchmark for
assessing the performance of other estimators \citep[see,
e.g.,][]{an+evans2011}.

\end{enumerate}

\noindent
The mean and standard deviation of the virial-theorem and 
orbital-roulette estimators, and of the generating-function estimator GF0 are
shown in Figure \ref{fig:three} over the range $N=5$--1000, for a
distribution of semimajor axes having $\gamma=0$ and
$a_{\rm max}/a_{\rm min}=3$. In these simulations the squared
eccentricity is distributed uniformly random between 0 and 1,
corresponding to a \df\ that is constant on an energy surface in
canonical phase space. At $N=1000$ the standard deviations in
$\hat\mu$, ordered by increasing magnitude, are: GF0, 0.013; orbital
roulette using the Anderson--Darling test, 0.018; orbital roulette
using the mean phase, 0.022; and virial theorem, 0.031. Thus the
performance of GF0 is substantially better than the other
estimators. Note that the GF0 estimator is independent of the
distribution of semimajor axes in the sample, and its superior
performance holds for a wide range of possible semimajor axis
distributions.

\begin{figure}
\centering
\includegraphics[width=\columnwidth]{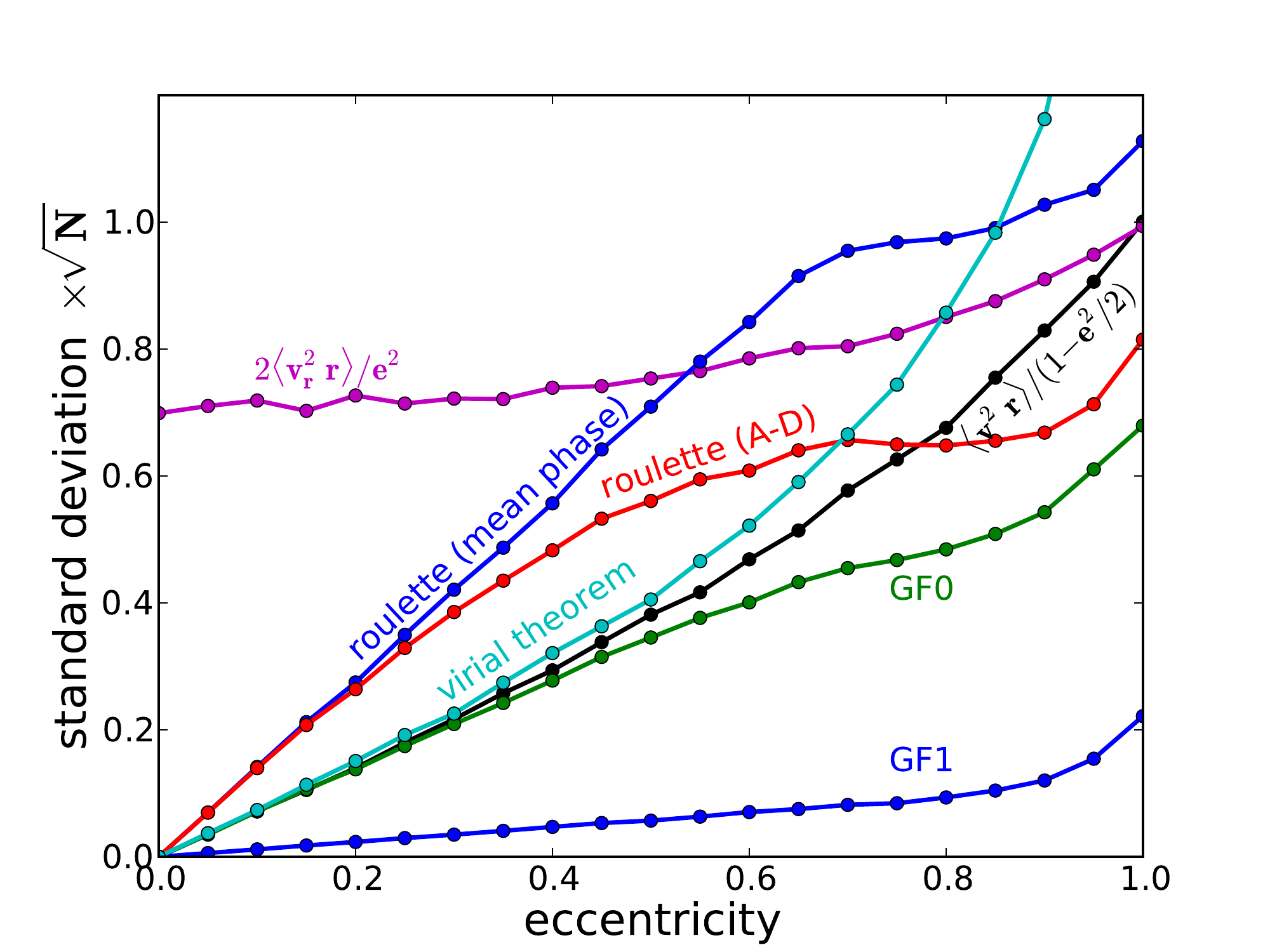}
\caption {\small Performance of estimators $\hat\mu$ for 100 particles orbiting
  in a Kepler potential, all with the same eccentricity (horizontal
  axis). The standard deviation of $\hat\mu$, multiplied by
  $\sqrt{100}$ to provide a quantity independent of the number of particles, is plotted on the
  vertical axis for each of the following: virial theorem (eq.\
  \ref{eq:virial-m}, cyan line); estimator based on the mean of
  $v^2r$ (eq.\ \ref{eq:meanx}, black line); estimator based on the
  mean of $v_r^2r$ (eq.\ \ref{eq:meany}, magenta line);  orbital
  roulette using the mean phase (blue line); orbital roulette using
  the Anderson--Darling test (red line); the generating-function estimator GF0 ($j_\star=0$;
  eq. \ref{eq:mkep0}, green line); and the generating-function
  estimator GF1 ($j_\star=j_\mathrm{min}$, blue line). The standard
  deviation is measured from 5000 realizations. The true mass is $\mu=1$, and the
  distribution of semimajor axes is given by equation
  (\ref{eq:dist-a}) with $\gamma=0$ and $a_{\rm max}/a_{\rm
    min}=3$. The results plotted near $e=0$ and $e=1$ are for $e=0.0001$
  and $e=0.9999$.}
\label{fig:four}
\end{figure}

The minimum-variance estimator GF1 does not perform significantly
better than GF0 in these simulations. The reason can be traced to
equation (\ref{eq:jmin-m}) for $j_\mathrm{min}$. In these simulations
we have assumed that the probability distribution of the squared
eccentricity is uniform, $dp =de^2$. Then for a given semimajor axis,
the mean of the denominator in the expression for $j_\mathrm{min}$ is
proportional to $\int_0^1 de\, e(1-e^2)^{1/2}[1-(1-e^2)^{1/2}]^{-1}$,
which diverges logarithmically at its lower limit. Thus in any large
sample $j_\mathrm{min}$ will be small, so the estimator GF1 will
perform similarly to GF0. On the other hand, for a sample of particles
with fixed non-zero eccentricity GF1 performs remarkably well. This is
illustrated in Figure \ref{fig:four}, which shows the normalized
standard deviation (i.e., the standard deviation times $\surd N$) of
several mass estimators as a function of the eccentricity. GF1 is much
more accurate than any of its competitors.

A final test, motivated by \cite{bovy+2012}, is to ask how well the
generating-function estimators would predict the solar mass, given a
snapshot of the positions and velocities of the eight planets in the
solar system. Given the current semimajor axes and eccentricities of
the planets, sampling the positions and velocities at a random time
and applying the estimator GF0 yields the correct solar mass with a
fractional standard deviation of 2.1 per cent. Using the 
positions and velocities on 2009 April 1 (taken from Table 1 of \citealt{bovy+2012})
yields $\hat\mu/(GM_\odot)=1.028$. Thus, GF0 provides mass estimates
accurate to within about 3 per cent in this case, even though it is given only 8 data
points and no prior information on the eccentricity or semimajor axis
distributions. 

\section{Discussion}

\noindent
These results offer encouraging evidence of the power of these
estimators, but unanswered questions remain.

Can the generating-function estimators be derived from a
maximum-likelihood approach? A possible signpost is that the GF0
estimator in a harmonic potential yields the maximum likelihood of the
frequency $\omega$ for the data set $\{ v_n/x_n\}$, the unique
combination of positions and velocities that has the same dimension as
$\omega$ (see discussion at the end of \S\ref{sec:maxl}). However, no
analogous result is known for most other potentials, including the Kepler
potential.

What is the relation of the generating-function approach to Bayesian
estimates of the posterior distribution of the mass parameters given
the data \citep[e.g.,][]{bovy+2012}?

We have shown that the generating-function estimators perform better than other
distribution-free estimators such as the virial theorem or orbital
roulette when applied to the harmonic or Kepler potentials. We have
also shown that they avoid the conceptual difficulties associated with
methods that estimate the distribution function simultaneously with
the mass parameters, such as Schwarzschild's method. We have not,
however, directly compared the efficiency of the latter methods to the
efficiency of generating-function estimators when applied to the same data.

In practical problems of mass estimation in astrophysics, the data are
often available only over a subset of the volume occupied by the
stellar system, either because of the limited field of view of the
survey or because of crowding near the core or background
contamination in the outskirts. How can generating-function estimators
be modified to account for both incomplete phase-space sampling and observational
errors?

A difficult problem arises when not all of the phase-space coordinates
are known, typically because the distance along the line of sight
cannot be determined accurately, and the proper motions are too small
to be detectable. Can the generating-function estimators be
generalized to the case where only some of the six phase-space
coordinates of the particle are known?

\section{Summary}

\noindent
We have described a novel method based on generating functions for
estimating the mass parameters of a gravitational potential, given
that we have positions and velocities for a set of test particles
orbiting in the potential in a steady state (i.e., with random
phases). The method we describe is (i) distribution-free, in the sense
that it requires no prior assumptions about the distribution of the
particles in action space; (ii) iterative, in the sense that it
produces an estimate $\hat\mu$ for the mass given a trial value
$\mu_t$, (iii) unbiased, in the sense of equation (\ref{eq:bias}). In
practice we iterate the estimate until $\hat\mu=\mu_t$ (eq.\
\ref{eq:conv}). We have demonstrated that this estimator is more
powerful than other distribution-free estimators in the
harmonic and Kepler potentials.

These results point the way towards more efficient and reliable mass
estimators that could have broad applications in astrophysical
dynamics.

\appendix

\section{Estimators for systems with multiple mass parameters and
  multiple degrees of freedom}

\subsection{Generalization to several mass parameters}

\noindent
The results of \S \ref{sec:gf} can be generalized to systems with
$M>1$ mass parameters $\bfmu=(\mu_1,\ldots,\mu_M)$ and one degree of freedom. The true but
unknown mass parameters are described by the vector $\bfmu$, and we
choose a trial vector $\bfmu_t$ that is close to $\bfmu$, with
$\Delta\bfmu\equiv \bfmu-\bfmu_t$. The estimator of $\bfmu$ is (cf.\
eq.\ \ref{eq:hatdef})
\begin{equation}
\hat\bfmu=\bfmu_t+\frac{1}{N}\sum_{n=1}^N\bfP(j_{t,n},\theta_{t,n},\bfmu_t)
\label{eq:w1}
\end{equation}
and we require that the estimator is consistent and unbiased up to
errors of $\mbox{O}(|\Delta\bfmu|^2)$. The analog to the
generating function in equation (\ref{eq:genfuna}) is
\begin{equation}
s(j,\theta_t,\bfmu_t,\Delta\bfmu)=\sum_{\alpha=1}^M \Delta\mu_\alpha
s_\alpha(j,\theta_t,\bfmu_t)+\mbox{O}(\Delta\mu)^2,
\label{eq:genfunb}
\end{equation}
and the condition on a component of the vector
$\bfP$ analogous to (\ref{eq:pppqqq}) is
\begin{align}
&\langle P_\alpha \rangle_\theta + \sum_{\beta=1}^M \Delta\mu_\beta\langle (\p_1 P_\alpha)(\p_2 s_\beta) - (\p_2 P_\alpha)(\p_1 s_\beta)
\rangle_\theta\nonumber \\
&= \Delta\mu_\alpha +\mbox{O}(\Delta\mu)^2;
\label{eq:ggg}
\end{align}
in this equation the arguments of all functions are $(j,\theta,\bfmu_t)$.
Let 
\begin{equation}
P_\alpha(j,\theta,\bfmu)=\sum_{\gamma=1}^M
c_{\alpha\gamma}(j,\bfmu)\p_2 s_\gamma(j,\theta,\bfmu);
\end{equation}
then
$\langle P_\alpha\rangle_\theta=0$ and equation (\ref{eq:ggg}) simplifies to
\begin{equation}
\sum_{\gamma=1}^M \frac{\p}{\p j} c_{\alpha\gamma}(j,\bfmu)
S_{\gamma\beta}(j,\bfmu)=\delta_{\alpha\beta}
\end{equation}
where
\begin{equation}
S_{\gamma\beta}(j,\bfmu)=\langle \p_2s_\gamma\,\p_2 s_\beta\rangle_\theta.
\label{eq:gggg}
\end{equation}
Now let $\matS^{-1}$ denote the inverse of the $M\times M$ matrix
$\matS$ with components $S_{\alpha\gamma}$.  If we set
$c_{\alpha\gamma}=(j-j_\star)\,S^{-1}_{\alpha\gamma}$ then equation
(\ref{eq:gggg}) is satisfied, and a suitable estimator is given by (\ref{eq:w1})
with 
\begin{equation}
P_\alpha=(j-j_\star)\sum_{\gamma=1}^M S^{-1}_{\alpha\gamma}\,\p_2
s_\gamma(j,\theta,\bfmu).
\end{equation}

\subsection{Generalization to several degrees of freedom}

\noindent
Now suppose that there is a single mass parameter $\mu$ and $D>1$
degrees of freedom, so the actions and angles become $D$-dimensional
vectors $(\bfj,\bftheta)$. The analog to equation (\ref{eq:pppqqq}) is
\begin{equation}
\langle P \rangle_\theta + \Delta\mu\sum_{k=1}^D \left\langle \frac{\p
    P}{\p j_k}\frac{\p s_1}{\p\theta_k} - \frac{\p  P}{\p
    \theta_k}\frac{\p s_1}{\p j_k} \right\rangle_{\!\bftheta} = \Delta\mu +\mbox{O}(\Delta\mu)^2;
\label{eq:pmulti}
\end{equation}
here the average $\langle\cdot\rangle_{\bftheta}$ is over all of the $D$
angles. Let us write 
\begin{equation}
P(\bfj,\bftheta,\mu)= \sum_i
q_i(\bfj,\bftheta,\mu)\, \p s_1(\bfj,\bftheta,\mu)/\p\theta_i.
\end{equation}
 Then equation (\ref{eq:pmulti}) requires
\begin{equation}
\sum_{k,i=1}^D \frac{\p}{\p j_k} T_{ki}(\bfj,\mu)q_i(\bfj,\mu)=1,
 \ \ \mbox{where}\ \ T_{ki}(\bfj,\mu)= \left\langle \frac{\p s_1}{\p\theta_k}\frac{\p s_1}{\p \theta_i} \right\rangle_{\!\bftheta}.
\label{eq:p2}
\end{equation}
This condition is satisfied if
\begin{equation}
q_i(\bfj,\mu)=\sum_{m=1}^D T^{-1}_{im}\alpha_m
(j_m-j_m^\star)\quad\mbox{where}\quad \sum_{m=1}^D\alpha_m=1.
\end{equation}
where $\matT^{-1}$ is the inverse of the matrix $\matT$ and
$\bfj^\star\}$ is a constant vector. 

\subsection{Generalization to several mass parameters and degrees of
  freedom}

\noindent
It is straightforward to generalize the preceding results to systems with $M>1$ mass parameters and $D>1$ degrees of
freedom. 
The analogs to equations (\ref{eq:w1})--(\ref{eq:ggg}) are
\begin{equation}
\hat\bfmu=\bfmu_t+\frac{1}{N}\sum_{n=1}^N\bfP(\bfj_{t,n},\bftheta_{t,n},\bfmu_t);
\label{eq:w1m}
\end{equation}
\begin{equation}
s(\bfj,\bftheta_t,\bfmu_t,\Delta\bfmu)=\sum_{\alpha=1}^M \Delta\mu_\alpha
s_\alpha(\bfj,\bftheta_t,\bfmu_t)+\mbox{O}(\Delta\mu)^2,
\label{eq:genfunm}
\end{equation}
\begin{align}
&\langle P_\alpha \rangle_\theta + \sum_{\beta=1}^M
\Delta\mu_\beta\sum_{k=1}^D\left\langle \frac{\p P_\alpha}{\p j_k}\frac{\p
  s_\beta}{\p\theta_k} - \frac{\p P_\alpha}{\p\theta_k}\frac{\p s_\beta}{\p j_k}
\right\rangle_{\bftheta} \nonumber \\
&= \Delta\mu_\alpha +\mbox{O}(\Delta\mu)^2.
\label{eq:gggm}
\end{align}
Here $P_\alpha$ is a component of an $M$-dimensional vector
$\bfP(\bfj,\bftheta,\bfmu)$. We choose this to have the form
\begin{equation}
P_\alpha(\bfj,\bftheta,\bfmu)=\sum_{\gamma=1}^M\sum_{i=1}^D Q_{\alpha
  i\gamma}(\bfj,\bftheta,\bfmu)\frac{\p s_\gamma(\bfj,\bftheta,\bfmu)}{\p\theta_i}.
\end{equation}
The condition (\ref{eq:gggm}) becomes 
\begin{equation}
\sum_{\gamma=1}^M\sum_{i,k=1}^D \frac{\p}{\p j_k }Q_{\alpha i\gamma
  }(\bfj,\bftheta,\bfmu) U_{i\gamma
    k\beta}(\bfj,\bftheta,\bfmu)=\delta_{\alpha\beta}
\end{equation}
where
\begin{equation}
U_{i\gamma k\beta}\equiv \left\langle \frac{\p s_\gamma}{\p
    \theta_i}\frac{\p s_\beta}{\p \theta_k}\right\rangle_{\bftheta}.
\label{eq:big}
\end{equation}

Now introduce a multi-index $\mathfrak{i}$ where each
$\mathfrak{i}$ corresponds to a 2-tuple $(i,\gamma)$. Similarly the
indices $\mathfrak{k}$ and $\mathfrak{m}$ correspond to the pairs
$(k,\beta)$ and $(m,\alpha)$ respectively. Then equation
(\ref{eq:big}) can be rewritten 
\begin{equation}
\sum_{k=1}^D \frac{\p}{\p j_k } \sum_{\mathfrak{i}\le(D,M)}Q_{\alpha\mathfrak{i}}(\bfj,\bftheta,\bfmu) U_\mathfrak{i
  k}(\bfj,\bftheta,\bfmu)=\delta_{\alpha\beta}.
\label{eq:biggg}
\end{equation}
If we set
\begin{equation}
Q_{\alpha\mathfrak{i}}(\bfj,\bftheta,\bfmu)\equiv \sum_{m=1}^D \alpha_m( j_m-j^\star_m)
U^{-1}_\mathfrak{m i}(\bfj,\bftheta,\bfmu)\ \ \mbox{where}\ \ \sum_{m=1}^D\alpha_m=1,
\end{equation}
then the left side of  equation (\ref{eq:biggg}) becomes
\begin{align}
&\sum_{k,m=1}^D \frac{\p}{\p j_k }\alpha_m
(j_m-j^\star_m)\sum_{\mathfrak{i}\le(D,M)}^{DM}U^{-1}_\mathfrak{m i}((\bfj,\bftheta,\bfmu) U_\mathfrak{i
  k}(\bfj,\bftheta,\bfmu)\nonumber \\
&=\sum_{k,m=1}^D \frac{\p}{\p j_k }
\alpha_m(j_m-j^\star_m)\delta_{\alpha\beta}\delta_{mk}=\delta_{\alpha\beta}
\label{eq:biggs}
\end{align}
so condition (\ref{eq:biggg}) is satisfied. 

% \bsp
\label{lastpage}

\begin{thebibliography}{99}


\bibitem[Aaronson(1983)]{aar1983} Aaronson, M.\ 1983, \apjl, 266, L11 

\bibitem[An \& Evans(2011)]{an+evans2011} An, J.~H., \& Evans, N.~W.\ 2011, \mnras, 413, 1744 

\bibitem[Bahcall \& Tremaine(1981)]{bt81} Bahcall, J.~N., \& Tremaine, S.\ 1981, \apj, 244, 805 

\bibitem[Beloborodov \& Levin(2004)]{bl04} Beloborodov, A.~M., \&
  Levin, Y.\ 2004, \apj, 613, 224 

\bibitem[Bovy et al.(2010)]{bovy+2012} Bovy, J., Murray, I., \& Hogg, D.~W.\ 2010, \apj, 711, 1157 

\bibitem[Chakrabarty \& Saha(2001)]{saha01} Chakrabarty, D., \& Saha, P.\ 2001, \aj, 122, 232 

\bibitem[Feldmeier-Krause et al.(2017)]{feld16} Feldmeier-Krause, A., Zhu, L., Neumayer, N., et al.\ 2017, \mnras, 466, 4040 

\bibitem[Kapteyn(1922)]{kap22} Kapteyn, J.~C.\ 1922, \apj, 55, 302 

\bibitem[Lancaster(2000)]{lan2000} Lancaster, T.\ 2000,  J.\ Econometrics, 95, 391

\bibitem[Magorrian(2006)]{mag06} Magorrian, J.\ 2006, \mnras, 373, 425 

\bibitem[Magorrian(2014)]{mag14} Magorrian, J.\ 2014, \mnras, 437, 2230 

\bibitem[McConnachie(2012)]{mac12} McConnachie, A.~W.\ 2012, \aj, 144, 4 

\bibitem[Neyman \& Scott(1948)]{ns48} Neyman, J., \& Scott, E.\ L.\
  1948, Econometrica 16, 1

\bibitem[Zwicky(1933)]{zw33} Zwicky, F.\ 1933, Helvetica Physica Acta, 6, 110 

\end{thebibliography}
\end{document}